\documentstyle{article}
\title{Time is dispensable in thermodynamics}
\author{Newton C. A. da Costa\\Research Group on Logic and Foundations.\\Institute for Advanced Studies, University of S\~{a}o Paulo.\\Av. Prof. Luciano Gualberto, trav. J, 374.\\05655-010 S\~{a}o Paulo SP Brazil. \and Adonai S. Sant'Anna\thanks{To whom correspondence should be sent. E-mail: adonai@scientist.com URL: http://www.geocities.com/adonaisantanna/adonaiss.html}\\Department of Mathematics, Federal University of Paran\'a\\C. P.
019081, 81531-990 Curitiba, PR, Brazil.}
\begin{document}
\maketitle
\newtheorem{definicao}{Definition}
\newtheorem{teorema}{Theorem}
\newtheorem{lema}{Lemma}
\newtheorem{corolario}{Corolary}
\newtheorem{proposicao}{Proposition}
\newtheorem{axioma}{Axiom}
\newtheorem{observacao}{Observation}

\begin{abstract}
We use Padoa's principle of independence of primitive symbols in axiomatic systems in order to show that time is dispensable in continuum thermodynamics, according to the axiomatic formulation of Gurtin and Williams. We also show how to define time by means of the remaining primitive concepts of Gurtin and Williams system. Finally, we introduce thermodynamics without time and briefly discuss some physical and philosophical consequences of our main results.
\end{abstract}

\tableofcontents

\section{Introduction}

	The present paper has a philosophical aspect in the sense that it copes with philosophical questions regarding the foundations of thermodynamics. On the other hand, it is also a work on mathematical 
physics, in the sense that we are concerned with the {\em mathematical\/} 
foundations of thermodynamics. Many branches of mathematics provide useful tools for theoretical physics, like probabilities, distributions, special functions, functional analysis, differential calculus, and so on. But few people use logic in theoretical physics. We consider this paper as a work on mathematical physics because we use mathematical logic in order to answer some fundamental questions concerning the mathematical role of time in thermodynamics. In a recent work \cite{daCosta-??} we have shown, by using Padoa's principle of independence of primitive concepts, that time is eliminable in Newtonian mechanics and that space-time is also dispensable in Hamiltonian mechanics, Maxwell's electromagnetic theory, the Dirac electron, classical gauge fields, and general relativity. Nevertheless, in all these theories physical phenomena are reversible with respect to time. Thermodynamics is a theory which talks about irreversible physical phenomena and so it seems to point out to an ``arrow of time''. There are many physical and philosophical discussions, in the liteature, about the objective existence of an arrow of time. For some authors there is, indeed, an objective ``passage'' of time \cite{Horwitz-88,Prigogine-80,Davies-95}. Nevertheless any concept of an objective passage of time seems to entail logical loops or to invoke ``absolute time'', which contradicts relativity theory \cite{Elitzur-99,Godel-49}. Other authors believe that most physicists prefer to regard time's passage as an illusion, since in theories like Einstein-Minkowski spacetime, all events (past, present, and future) have the same level of reality \cite{Elitzur-99}. In \cite{Prigogine-80} (p. 203) Prigogine makes reference to a private letter written by A. Einstein where it is written:

\begin{quote}
For us who are convinced physicists, the distinction between past, present, and future is only an illusion, however persistent.
\end{quote}

	Prigogine himself (op. cit. p. 213) says that

\begin{quote}
The distinction between past and future is a kind of {\em primitive concept\/} that in a sense precedes scientific activity.
\end{quote}

	But some paragraphs later he remarks that classical science is trying ``to go beyond the world of appearances, to reach a timeless world of supreme rationality.''

	In this paper we study the role of time in thermodynamics from the mathematical point of view. Since the theory is supposed to give a picture of physical phenomena, we hope that our results may be useful for the understanding of the physical and philosophical meaning of time in a theory which admits the existence of irreversible phenomena.

	Some decades ago M. E. Gurtin and W. O. Williams presented an axiomatic framework for continuum thermodynamics \cite{Gurtin-67}. In the present paper we rewrite Gurtin and Williams axiomatic system as a Bourbaki species of structures and prove that time is definable, and so is dispensable, in thermodynamics.

	For a brief review on the use of the axiomatic method in physics as well as other proposals for the interpretation of time and spacetime in physics see \cite{daCosta-??}.

\section{Padoa's Principle}

	This section is essentially based on our previous work \cite{daCosta-??}.

	In an axiomatic system $S$ a primitive term or concept $c$ is definable by means of the remaining primitive ones if there is an appropriate formula, provable in the system, that fixes the meaning of $c$ in function of the other primitive terms of $S$. This formulation of definability is not rigorous but is enough here. When $c$ is not definable in $S$, it is said to be independent of the the other primitive terms.

	There is a method, introduced by A. Padoa \cite{Padoa-00}, which can be employed to show the independence of concepts. In fact, Padoa's method gives a necessary and sufficient condition for independence \cite{Beth-53,Suppes-57,Tarski-83}.

	In order to present Padoa's method, some preliminary remarks are necessary. Loosely speaking, if we are working in set theory, as our basic theory, an axiomatic system $S$ characterizes a species of mathematical structures in the sense of Bourbaki \cite{Bourbaki-68}. Actually there is a close relationship between Bourbaki's species of structures and Suppes predicates \cite{Suppes-67}; for details see \cite{daCosta-88}. On the other hand, if our underlying logic is higher-order logic (type theory), $S$ determines a usual higher-order structure \cite{Carnap-58}. In the first case, our language is the first order language of set theory, and, in the second, it is the language of (some) type theory. Tarski showed that Padoa's method is valid in the second case \cite{Tarski-83}, and Beth that it is applicable in the first \cite{Beth-53}.

	From the point of view of applications of the axiomatic method, for example in the foundations of physics, it is easier to assume that our mathematical systems and structures are contructed in set theory \cite{daCosta-88}.

	A simplified and sufficiently rigorous formulation of the method, adapted to our exposition, is described in the next paragraphs.

	Let $S$ be an axiomatic system whose primitive concepts are $c_1$, $c_2$, ..., $c_n$. One of these concepts, say $c_i$, is independent from the remaining if and only if there are two models of $S$ in which $c_1$, ..., $c_{i-1}$, $c_{i+1}$, ..., $c_n$ have the same interpretation, but the interpretations of $c_i$ in such models are different.

	Of course a model of $S$ is a set-theoretical structure in which all axioms of $S$ are true, according to the interpretation of its primitive terms \cite{Mendelson-97}.

	It is important to recall that, according to the theory of definition \cite{Suppes-57}, a definition should satisfy the {\em criterion of eliminability\/}. That means that a defined symbol should always be eliminable from any formula of the theory. 

	In the sequel we apply Padoa's method to thermodynamics (in an axiomatic form), in order to prove that time is eliminable (or dispensable).

\section{Gurtin and Williams Axiomatic System}

	In this section we present the axiomatic system of thermodynamics due to Gurtin and Williams \cite{Gurtin-67}. We present it as a Bourbaki's species of structure. This demands some technical adaptations which do not affect the physical meaning of the theory. But such adaptations are necessary in order to use Padoa's method. We use the same notation as in \cite{Gurtin-67}.

	It is interesting to settle our mathematical notation. We denote by ${\cal E}$ the Euclidian three dimensional space. $\Re$ is the set of real numbers. If ${\cal A}$ is a subset of ${\cal E}$ then its boundary is denoted by $\partial {\cal A}$, the interior by $\stackrel{\;\;\circ}{\cal A}$, and the closure by $\overline{\cal A}$. The letters ${\cal A}$, ${\cal B}$, ${\cal C}$, ..., always denote subsets of ${\cal E}$.

\begin{definicao}
A {\em standard region\/} or {\em body\/} of ${\cal E}$ is the closure of a bounded open set ${\cal A}$ whose boundary is the union of a closed set of zero area and a countable number of class $C^1$ two-dimensional manifolds, each of the manifolds having the open set ${\cal A}$ on just one side.
\end{definicao}

\begin{definicao}
If ${\cal A}$ is a subset of ${\cal E}$, its {\em exterior\/} is 

$${\cal A}^e = \overline{{\cal E} - {\cal A}}$$
\end{definicao}

\begin{definicao}
If ${\cal A}\subset {\cal B}$, the {\em relative exterior\/} of ${\cal A}$ in ${\cal B}$ is

$${\cal A}^b = \overline{{\cal B} - {\cal A}}.$$
\end{definicao}

\begin{teorema}
$${\cal A}^e = {\cal A}^b \cup {\cal B}^e.$$
\end{teorema}

\begin{description}
\item [Proof:] Straightforward from definitions.
\end{description}

\begin{definicao}
A {\em surface\/} is the (relative) closure of an oriented class $C^1$ two-dimensional differentiable manifold or the countable union of such (closed) manifolds.
\end{definicao}

	According to \cite{Gurtin-67}: ``The boundary of a standard region is taken to be oriented in the positive sense with respect to that region, i.e. with the orientation corresponding to the external normal vector. A surface contained (in the sense of set-inclusion) in another surface is a {\em positive segment\/} of that surface if it has the same orientation; if it has the opposite orientation it is called a {\em negative segment\/}.''

\begin{definicao}
A surface ${\cal S}$ contained in a body ${\cal B}$ is a {\em material surface\/} if it is a positive segment of the boundary of a subbody of ${\cal B}$.
\end{definicao}

\begin{definicao}
A {\em part\/} of a body ${\cal B}$ is a Borel subset of ${\cal B}$. The set of Borel subsets of ${\cal B}$ is denoted by ${\bf B}({\cal B})$.
\end{definicao}

	By {\em measure\/} we mean a finite real-valued Borel signed measure \cite{Doob-94}.

\begin{definicao}
The species of structures (\`a la Bourbaki) of a {\em Gurtin-Williams System for Continuum Thermodynamics\/} is the ordered $6$-tuple

$$\Theta = \langle {\cal E}, T, E_t, H_t, S_t, M_t \rangle$$

\noindent
such that the following axioms are satisfied:

\begin{description}

\item [T1 - Space - ] ${\cal E}$ is the Euclidean three dimensional space.

\item [T2 - Subbodies of a Body - ] For every body ${\cal B}$ there is a class of subbodies ${\cal M}^{\cal B}$ such that 

\begin{enumerate}

\item Every element of ${\cal M}^{\cal B}$ is a subset of ${\cal B}$;

\item Every element of ${\cal M}^{\cal B}$ is a body;

\item ${\cal A}, {\cal C} \in {\cal M}$ implies ${\cal A}\cup {\cal C} \in {\cal M}^{\cal B}$;

\item If ${\cal C}$ is a solid circular cylinder or a solid prism in ${\cal E}$, then

$$\overline{\stackrel{\circ}{\overbrace{{\cal C}\cap {\cal B}}}} \in {\cal M}^{\cal B};$$

\item If ${\cal S}\subset {\cal B}$ is a material surface, there is a monotone sequence $\{ {\cal A}_n\}$ of elements of ${\cal M}^{\cal B}$ such that 

$$\bigcap_{n=1}^\infty {\cal A}_n = {\cal S};$$

\item If ${\cal A} \in {\cal M}^{\cal B}$ and ${\bf a}$ is a vector in ${\cal E}$, then ${\cal A}+{\bf a} \in {\cal M}^{\cal B}$.

\end{enumerate}

\item [T3 - Time - ] $T$ is an interval of real numbers, which we interpret as {\em time\/}.

\item [T4 - Internal Energy - ] For each body ${\cal B}$ there is an {\em energy function\/} $E_t: {\bf B}({\cal B})\times T \to \Re$ such that $E_t$ is a measure on ${\bf B}({\cal B})$. This function corresponds to the internal energy of each part of the body ${\cal B}$.

\item [T5 - Differentiability of Energy - ] For each body ${\cal B}$ and for each part ${\cal P}\in {\bf B}({\cal B})$ there exists the derivative

$$\stackrel{\bullet}{E}_t({\cal P}) = \frac{d}{dt}E_t({\cal P}).$$

\item [T6 - Volume and Energy - ] For all ${\cal P}\in {\bf B}({\cal B})$ there exists a scalar $\alpha(t)$ such that 

$$|E_t({\cal P})|\leq \alpha(t) V({\cal P}),$$

\noindent
and

$$|\stackrel{\bullet}{E}_t({\cal P})|\leq \alpha(t) V({\cal P}),$$

\noindent
where $V$ is the Lebesgue volume measure in ${\cal E}$.

\end{description}

\begin{definicao}
The {\em material universe\/} for a given body ${\cal B}$ is the set

$${\cal M} = \{ {\cal D}| {\cal D} \in {\cal M}^{\cal B} \mbox{or} \;{\cal D}^e \in {\cal M}^{\cal B}\}.$$

\end{definicao}

\begin{description}

\item [T7 - Heat Flux - ] For each body ${\cal B}$ of ${\cal E}$ and for each element ${\cal D}$ of the material universe for ${\cal B}$ there is a {\em heat flux\/} function $H_t: {\bf B}({\cal D}^b)\times {\cal M} \times T \to \Re$ such that $H_t$ is a measure on ${\bf B}({\cal D}^b)$ for a fixed ${\cal D}$.

\end{description}

\begin{definicao}
A real-valued function $\alpha$ defined on ${\cal M}^{\cal B}$ or ${\cal M}$ is {\em separately additive\/}, or {\em s-additive\/} for short, if

$$\alpha({\cal A}\cup{\cal C}) = \alpha({\cal A}) + \alpha({\cal C})$$

\noindent
for every pair of separate elements ${\cal A}$, ${\cal C}$ in the domain of $\alpha$.
\end{definicao}

\begin{description}

\item [T8 - Heat Flux is Separately Additive - ] For a fixed part ${\cal P}$ of a given body ${\cal B}$ the heat flux $H_t({\cal P}, \bullet)$ is s-additive on all elements of ${\cal M}$ separate from ${\cal P}$.

\item [T9 - Surface and Heat Flux- ] There exist scalar functions $\beta(t)$ and $\gamma(t)$ such that 

$$|H_t({\cal P},{\cal D})|\leq \beta(t)V({\cal P}) + \gamma(t) A({\cal P}\cap \partial {\cal D})$$

\noindent
for all ${\cal P}\in {\bf B}({\cal B})$, ${\cal D}\in {\cal M}$ which are separate, where $A$ stands for the Lebesgue surface measure on manifolds in ${\cal E}$.

\item [T10 - First Law of Thermodynamics - ] For every body ${\cal A}$ 

$$\stackrel{\bullet}{E}_t({\cal A}) = H_t({\cal A},{\cal A}^e).$$

\item [T11 - Internal Entropy - ] For each body ${\cal B}$ there is an {\em internal entropy\/} function $S_t: {\bf B}({\cal B})\times T \to \Re$ such that $S_t$ is a measure on ${\bf B}({\cal B})$.

\item [T12 - Differentiability of Entropy - ] The derivative

$$\stackrel{\bullet}{S}_t({\cal P}) = \frac{d}{dt}S_t({\cal P})$$

\noindent
exists for each ${\cal P}\in {\bf B}({\cal B})$.

\item [T13 - Volume and Entropy - ] For all ${\cal P}\in {\bf B}({\cal B})$ there exists a scalar $\delta(t)$ such that 

$$|S_t({\cal P})|\leq \delta(t) V({\cal P}),$$

\noindent
and

$$|\stackrel{\bullet}{S}_t({\cal P})|\leq \delta(t) V({\cal P}),$$

\noindent
where, like in axiom {\bf T6}, $V$ is the Lebesgue volume measure in ${\cal E}$.

\item [T14 - Entropy Flux - ] For each body ${\cal B}$ of ${\cal E}$ and for each element ${\cal D}$ of the material universe for ${\cal B}$ there is an {\em entropy flux\/} function $M_t: {\bf B}({\cal D}^b)\times {\cal M} \times T \to \Re$ such that $M_t$ is a measure on ${\bf B}({\cal D}^b)$ for a fixed ${\cal D}$.

\item [T15 - Entropy Flux is Separately Additive - ] For a fixed part ${\cal P}$ of a given body ${\cal B}$ the entropy flux $M_t({\cal P}, \bullet)$ is s-additive on all elements of ${\cal M}$ separate from ${\cal P}$.

\end{description}

\begin{definicao}
A part ${\cal P}$ of a body ${\cal B}$ is {\em thermally isolated from \/} ${\cal D}\in {\cal M}$ if for each part ${\cal P}'\subset {\cal P}$

$$H_t({\cal P}', {\cal D}) = 0.$$
\end{definicao}

\begin{description}

\item [T16 - Second Law of Thermodynamics - ]

\begin{enumerate}

\item For every subbody ${\cal A}$ of a given body ${\cal B}$

$$\stackrel{\bullet}{S}_t({\cal A})\geq M_t({\cal A},{\cal A}^e);$$

\item If a part ${\cal P}$ is thermally isolated from ${\cal D}\in {\cal M}$

$$M_t({\cal P}, {\cal D}) = 0.$$

\end{enumerate}

\item [T17 - Heat Conduction Inequalities - ] There exist scalars $\delta(t)$ and $\varepsilon(t)$ such that

$$|K_t({\cal P},{\cal D})|\leq\delta(t)V({\cal P}),$$

$$|J_t({\cal P},{\cal D})|\leq\varepsilon(t)A({\cal P}\cap\partial{\cal D}),$$

\noindent
for all ${\cal D}\in {\cal M}$ and ${\cal P}\in {\bf B}({\cal D}^b)$, where 

$$K_t({\cal P},{\cal D}) = M_t({\cal P} - \partial {\cal D}, {\cal D})$$

\noindent
is the {\em radiative entropy flux\/} and

$$J_t({\cal P},{\cal D}) = M_t({\cal P}\cap\partial {\cal D}, {\cal D})$$

\noindent
is the {\em conductive entropy flux\/}.

\end{description}

\end{definicao}

	It is worth to remark that in \cite{Gurtin-67} there is a theorem that says that for any ${\cal D}\in {\cal M}$, $M_t$ admits the unique decomposition

$$M_t({\cal P}, {\cal D}) = K_t({\cal P},{\cal D}) + J_t({\cal P},{\cal D}).$$

	A very detailed analysis on these axioms is made in \cite{Gurtin-67}. At the present paper we are mainly concerned with the results introduced in the next section.

\section{Thermodynamics Without Time}

	In this section we: (i) prove that time is definable; (ii) define time from the remaining primitive concepts of Gurtin-Williams system for thermodynamics; and (iii) rephrase thermodynamics without time. So, our starting point is the next theorem:

\begin{teorema}
Time is dispensable in Gurtin-Williams system for continuum thermodynamics.
\end{teorema}

\noindent
{\it Proof:} Padoa's Principle says that the primitive concept $T$ in $\Theta$ is independent from the remaining primitive concepts iff there are two models of $\Theta$ such that $T$ has two interpretations and all the other primitive symbols have the same interpretation. But these two interpretations are not possible, since all the remaining concepts, except ${\cal E}$, depend on time $T$ as functions. Any change of interpretation related to $T$ will imply a change of interpretation of $E_t, H_t, S_t$, and $M_t$. Therefore, time is not independent and hence it can be defined. So, according to the criterion of eliminability of definitions, time is dispensable within the scope of this axiomatic framework for thermodynamics.$\Box$\\

	Since time is definable in continuum thermodynamics, the natural question is: how to define it? From the logical point of view the answer to this question is not difficult. But first we need the next definition:

\begin{definicao}
If the domain of a function $f$ is $A_1\times A_2\times \cdot A_n$, then we call each $A_i$ a {\em component\/} of the domain of $f$. For short, we say that $A_i$ is a component of $f$.
\end{definicao}

	Now we are able to define time in Gurtin-Williams system for thermodynamics:

\begin{definicao}
Time is the last component of the functions $E_t$, $H_t$, $S_t$, and $M_t$ and is the domain of the scalar functions $\alpha$, $\beta$, $\gamma$, and $\delta$.
\end{definicao}

	Now, the final question is: how to rephrase thermodynamics with no explicit mention to time? Our answer is the definition given below. Some axioms of this `new' system are the same as in $\Theta$, mainly those axioms that make no reference to time. We write the word `new' because actually it is not a new system. It is the same theory, but with no explicit mention to a definable (eliminable) term called `time'.

\begin{definicao}
The species of structures (\`a la Bourbaki) of a {\em Gurtin-Williams System for Continuum Thermodynamics Without Time\/} is the ordered $5$-tuple

$$\Theta_{NT} = \langle {\cal E}, E_t, H_t, S_t, M_t \rangle$$

\noindent
such that the following axioms are satisfied:

\begin{description}

\item [NT1] {\bf T1}.

\item [NT2] {\bf T2}.

\item [NT3] For each body ${\cal B}$ there is an {\em energy function\/} $E_t$ with two components. The first component is ${\bf B}({\cal B})$, such that $E_t$ is a measure on ${\bf B}({\cal B})$. The co-domain of this function is $\Re$.

\item [NT4] The function $E_t$ is differentiable with respect to the elements of the last (second) component of $E_t$. We denote the derivative as

$$\stackrel{\bullet}{E}_t({\cal P}) = \frac{d}{dt}E_t({\cal P}).$$

	This same derivative operator is also defined with respect to the last (second) component of $S_t$ on the space of functions $S_t$. We denote it by 

$$\stackrel{\bullet}{S}_t({\cal P}) = \frac{d}{dt}S_t({\cal P}).$$

\item [NT5] {\bf T6}.

\item [NT6] For each body ${\cal B}$ of ${\cal E}$ and for each element ${\cal D}$ of the material universe for ${\cal B}$ there is a function $H_t$ such that $H_t$ has three components. The first component is ${\bf B}({\cal D}^b)$, and the second component is the material universe ${\cal M}$. $H_t$ is a measure on ${\bf B}({\cal D}^b)$ for a fixed ${\cal D}$. The co-domain of the function $H_t$ is $\Re$.

\item [NT7] {\bf T8}.

\item [NT8] {\bf T9}.

\item [NT9] {\bf T10}.

\item [NT10] For each body ${\cal B}$ there is a function $S_t$ with two components. The first one is ${\bf B}({\cal B})$. $S_t$ is a measure on ${\bf B}({\cal B})$. The co-domain of $S_t$ is $\Re$.

\item [NT11] {\bf T13}.

\item [NT12] For each body ${\cal B}$ of ${\cal E}$ and for each element ${\cal D}$ of the material universe for ${\cal B}$ there is a function $M_t$ with three components. The first component is ${\bf B}({\cal D}^b)$ and the second is ${\cal M}$. $M_t$ is a measure on ${\bf B}({\cal D}^b)$ for a fixed ${\cal D}$. The co-domain of $M_t$ is $\Re$.

\item [NT13] {\bf T15}.

\item [NT14] {\bf T16}.

\item [NT15] {\bf T17}.

\end{description}

\end{definicao}

	It seems to be clear that this picture for thermodynamics is not very operational. So, for all practical purposes, it is still interesting to keep the notion of time. The eliminability of time should be rather regarded as a logical consequence from the foundations of thermodynamics.

\section{Final Remarks}

	In this section we make some final remarks concerning our main results:

\begin{enumerate}

\item The reader can verify that time is also eliminable from continuum mechanics, at least within the scope of the axiomatic system presented by W. Noll in \cite{Noll-59}. In that system some primitive notions like {\em motion\/} $\theta$, {\em linear momentum\/} ${\bf g}$ , and {\em angular momentum\/} ${\bf h}$ of a body are functions with time $T$ as one component. So, following Padoa's principle, any change of interpretation on $T$ entails a change of interpretation on other primitive concepts, like $\theta$, ${\bf g}$, and ${\bf h}$.

\item In \cite{daCosta-??} we did not define space-time in classical field theories or time in classical particle mechanics. Nevertheless, the same ideas presented in the previous section of the present paper may be used for that purpose. We keep this as an exercise for the reader.

\item We can easily see that space is also definable in Gurtin-Williams system. That is ledt as an exercise for the reader.

\item Some neurologists point out that blind people may have great difficulty to develop the notion of physical space \cite{Sacks-96}. They live in a spaceless world where time is their main reference frame. It seems that the vision in healthy people helps them to create an intuitive notion of space, since the visual perception allows people to be aware of many different objects and places at the same time. On the other hand, U. Mohrhoff \cite{Mohrhoff-00} presents a novel interpretation of quantum mechanics, where objective probabilities are assigned to counterfactuals and are calculated on the basis of all relevant facts, including those that are still in the {\em future\/}. According to this proposal, the intuitive distinction between here and there, past and future, has nothing to do with any physical reality. His starting point is a paper by N. D. Mermin \cite{Mermin-98}, where this author considers that conscious perception ``should be viewed as a mystery about {\em us\/} and should not be confused with the problem of understanding quantum mechanics.'' In the present paper we are not concerned with quantum mechanics. But we believe that the definability of time (and even space) could be interpreted as it follows: from our physical experience with the world, and our conscious perception of it, we develop some sort of spacetime interpretation about what we see and feel with all our senses. Much more should be said about our conscious perception of space and time. But that is a task that we leave for future works. In the present paper we want to point out the mathematical role of time in thermodynamics.

\end{enumerate}

\end{document}